\DocumentMetadata{}
\documentclass[sigconf,authorversion,preprint]{acmart}
\AtBeginDocument{%
  \providecommand\BibTeX{{%
    \normalfont B\kern-0.5em{\scshape i\kern-0.25em b}\kern-0.8em\TeX}}}

\copyrightyear{2024}
\acmYear{2024}
\setcopyright{acmlicensed}\acmConference[ICMI '24]{International Conference on Multimodal Interaction}{November 4--8, 2024}{San Jose, Costa Rica}
\acmBooktitle{International Conference on Multimodal Interaction (ICMI '24), November 4--8, 2024, San Jose, Costa Rica}
\acmDOI{10.1145/3678957.3685755}
\acmISBN{979-8-4007-0462-8/24/11}

\usepackage{xcolor}

\usepackage{multirow}
\usepackage{subcaption}
\usepackage{tabularx}
\usepackage{siunitx}

\usepackage{threeparttable}
\usepackage{CJKutf8}
\usepackage{pifont}
\colorlet{usercolorname}{red!0}

\begin{document}

\title[Juicy Text]{Juicy Text: Onomatopoeia and Semantic Text Effects \\ for Juicy Player Experiences}

\author{\'Emilie Fabre}
\affiliation{%
  \institution{The University of Tokyo}
  \city{Tokyo}
  \country{Japan}
}
\email{fabre@g.ecc.u-tokyo.ac.jp}
\orcid{0000-0003-1978-9374}

\author{Katie Seaborn}
\affiliation{%
  \institution{Institute of Science Tokyo}
  \city{Tokyo}
  \country{Japan}}
\email{seaborn@iee.eng.isct.ac.jp}
\orcid{0000-0002-7812-9096}

\author{Adrien Verhulst}

\affiliation{%
  \institution{Sony Computer Science Laboratories}
  \city{Tokyo}
  \country{Japan}}
\email{adrienverhulst @csl.sony.co.jp}
\orcid{0000-0003-4089-9025}

\author{Yuta Itoh}
\affiliation{%
  \institution{The University of Tokyo}
  \city{Tokyo}
  \country{Japan}
}
\email{yuta.itoh@iii.u-tokyo.ac.jp}
\orcid{0000-0002-5901-797X}

\author{Jun Rekimoto}
\affiliation{%
  \institution{The University of Tokyo}
  \city{Tokyo}
\country{Japan}}
\email{rekimoto@acm.org}
\orcid{0000-0002-3629-2514}

\renewcommand{\shortauthors}{Fabre et al.}

\begin{abstract}
Juiciness is visual pizzazz used to improve player experience and engagement in games. Most research has focused on juicy particle effects.
However, text effects are also commonly used in games, albeit not always juiced up. One type is onomatopoeia, a well-defined element of human language that has been translated to visual media, such as comic books and games.
Another is semantic text, often used to provide performance feedback in games. In this work, we explored the relationship between juiciness and text effects, aiming to replicate juicy user experiences with text-based juice and combining particle and text juice.
We show in a multi-phase within-subjects experiment that users rate juicy text effects similarly to particles effects, with comparable performance, and more reliable feedback. We also hint at potential improvement in user experience when both are combined, and how text stimuli may be perceived differently than other visual ones.
We contribute empirical findings on the juicy-text connection in the context of visual effects for interactive media.
\end{abstract}



\begin{teaserfigure}
    \centering
  \includegraphics[width=0.8\textwidth]{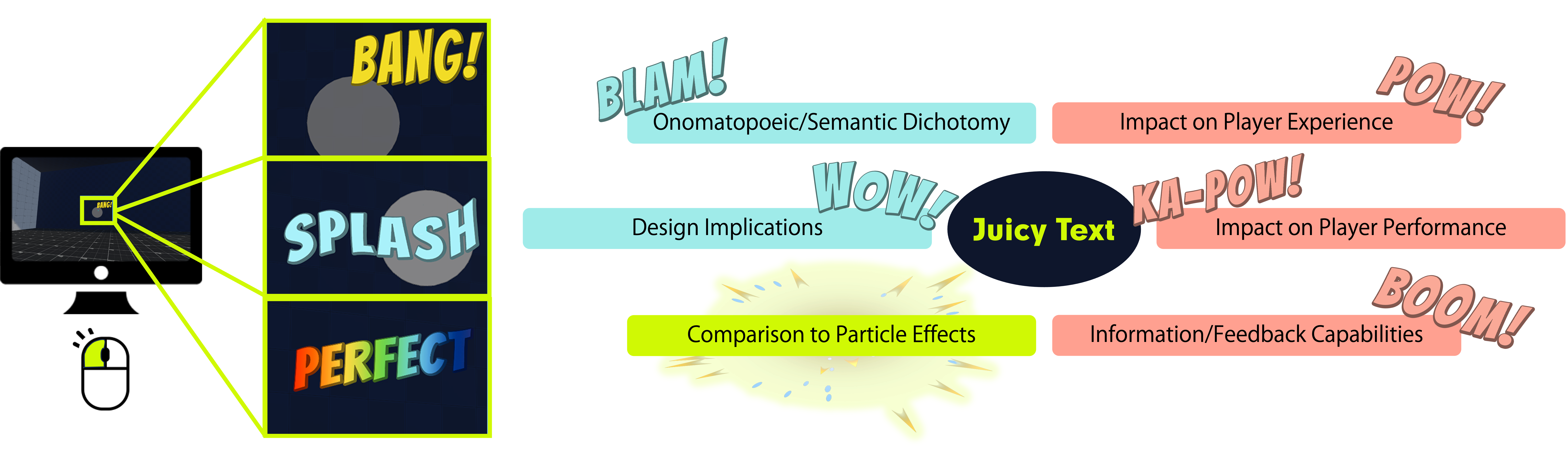}
  \caption{Left: Text effect samples used in the study. Right: Diagram of Juicy text concepts explored in this study.}
  \Description{Left: Samples of text effects used in this study, with a small diagram of a computer screen and a mouse to explain that the effects appear when clicking on the target. Right: Diagram of concepts surrounding Juicy text explored in this study. In the middle: Juicy Text. Around it we see Onomatopoeic/Semantic Dichotomy, Impact on Player Experience, Design Implications, Impact on Player Performance, Comparison to Particle Effects and Information/Feedback Capabilities}
  \label{fig:teaser}
\end{teaserfigure}

\maketitle


\section{Introduction}

The advancement of video games has driven the need for increasingly impressive Visual Effects (VFX). 
Players expect visually stunning experiences that immerse them in virtual worlds. This has driven innovation in VFX design~\cite{nordberg2020visual, Grissom_2018, Chamberlain_2017}.
A new concept has emerged that captures the essence of engaging VFX: \emph{Juiciness}~\cite{gray2005gamasutra}. Juiciness refers to the quality of VFX that creates impactful, responsive, and immersive player interactions. It encompasses the satisfaction of button presses, explosive feedback, and lively reactions. Incorporating juiciness into VFX design is now considered an important part of delivering captivating experiences~\cite{Pichlmair2021,Hicks2019}.
Recent research also explores juiciness' effects~\cite{Hicks2019,hicks2019gamification,Kao2020,juicyhaptic}, a design term, coined in 2005 by \citet{gray2005gamasutra}, presented as their "\textit{wet little term for constant and bountiful user feedback}" (paragraph 5). 
Yet, despite being grounded in player feedback and "game feel"~\cite{Pichlmair2021}, the notion of juiciness lacks clarity. At present, two common operationalizations exist in the literature: (i) \emph{amplification of the feedback needed by the player to understand the mechanics}~\cite[p. 19]{Pichlmair2021} and (ii) \textit{an abundance of audiovisual feedback}~\cite[p. 2]{hicks2019gamification}. 

Juiciness 
as a design tool enriches player experience (PX) through engaging visual elements like particles and animations~\cite{Pichlmair2021,juul2016good,Hicks2019,Kao2020,hastingsParticle2}, but also through audio or haptic effects~\cite{juicyhaptic,Pichlmair2021}. While textual effects are less explored, recent studies suggest their potential in enhancing PX by conveying semantic information~\cite{Oh2018,Wang2017,Fabre2021,kapow}. However, the impact and elements of juicy text (JT) remain unclear.

Onomatopoeia, which represents sound and action visually~\cite{Fabre2021,guynesOno,Sasamoto2019_showing}, is akin to juiced-up text. Combining juiciness and onomatopoeia could further enhance PX, although this remains unexplored. Additionally, the distinction between onomatopoeic and non-onomatopoeic text effects, as well as the potential of gibberish text, requires investigation.

We examined the impact of semantic and non-semantic JT effects on PX in interactive media. Specifically, we explored \textbf{how semantic JT, particularly onomatopoeic semantic text, influences PX and performance compared to non-semantic text and particle juice.} Through a three-part experiment, it reveals that while text alone may not significantly enhance a game, it can complement visual juicy effects.
The main contributions of this research are:

\begin{itemize}
    \item First known research on juicy text effects during play.
    \item Empirical evidence that onomatopoeic juicy text, combined with particles, enhances PX without performance degradation compared to a control (no juice effects).
    \item Empirical evidence that text content, compared to particle effects, is crucial for semantic feedback but not overall PX.
    \item Limited replication of findings on (i) juicy over standard particle effects and (ii) text conveyance of material information.
\end{itemize}

\section{Related Work}

\subsection{Modalities of Onomatopoeia and 
Text}
\label{sec:relworkono}
Onomatopoeia has the potential to link emotional experience and language~\cite{Sasamoto2019_showing}. 
\citet{Sasamoto2019_showing} argues that ``onomatopoeia is a bridge between verbal and non-verbal elements in multimodal media''~\cite[p. 28]{Sasamoto2019Intro}. This definition is relevant to game design, an emotion-driven medium~\cite{Pichlmair2021, okon2015neurobiology}.
Sasamoto et al.~\cite{Sasamoto2016} emphasize the strong communicative power of onomatopoeia with 
its dual role as "showing" and "saying"~\cite{wharton2003interjections} making it intriguing from a design perspective, as it enables the sharing of sensory experiences multimodally or cross-modally~\cite{Sasamoto2016}. It is difficult to map onomatopoeic words to actual words, leaving the sound represented by the onomatopoeic text as the sole conveyor of meaning~\cite{petersen2009acoustics}. 
This accounts for the popularity of onomatopoeia in manga and comic books~\cite{guynesOno,rohan2018argumentation}, where it assists writers in ``describing all five senses using only one
''~\cite[p. 146]{mccloud2006making}.
Indeed, onomatopoeia are commonly 
used to represent diegetic sounds occurring within the context of the story~\cite{guynesOno}. They serve as a visual soundtrack, termed "the comic book soundtrack" and "visual sound effects"~\cite{khordoc2001comic}. Especially popular in manga, or Japanese comic books, the stylized presentation of onomatopoeia increases its communicative power~\cite{Sasamoto2016,rohan2018argumentation}. This feature could be amplified in video games through design elements, such as colors and animation, and diegetic player and non-player action that occurs in response to player choices. \citet{Sasamoto2016} suggests that onomatopoeia function as "mimetics," or a form of mimicry, that is meant to break realism and surpass 
our expectations of reality~\cite{kusamori1968sutoori, natsume2013manga}. This maps onto the concept of juiciness~\cite{Pichlmair2021,Hicks2019}.

Onomatopoeia are especially prevalent in Japanese and Korean, often used to denote psychological states unrelated to sound~\cite{Sasamoto2019Intro,flyxe2002translation}. For our purpose, we define onomatopoeia as words directly mimicking or referencing sounds related to actions or events, like "boom," "splash," "zap," etc.

\subsection{Modalities of Juiciness and Juicy Player Experience: From Particles to Text}
“Juiciness”, coined by Kyle Gray in 2005~\cite{gray2005gamasutra}, became popular in game design and academia.
It has been studied for its potential to enhance PX by improving perceptions of competence and aesthetic appeal~\cite{Hicks2019,juul2016good}.
Different levels of juiciness may evoke distinct forms of PX, with excessive juiciness potentially degrading it~\cite{Kao2020}. This concept aligns with Swink's notion of "polish" in the theory of game feel~\cite{SwinkGameFeel}. This link was recently further solidified, as both emphasize aesthetic enhancement without altering core game mechanics~\cite{Pichlmair2021}. This has lead some to propose that "juicy" studies should compare the same underlying feedback, only with "unnecessary additional elements" in the juicy condition~\cite[p. 5]{tornqvist} or "\textit{how} feedback is presented"
~\cite[p. 186]{Hicks2019}. We use a minimal design, varying only type and level of juiciness. 

The use of juiciness in games often lacks clear design guidelines, with high-level recommendations like "make it juicy" prevailing~\cite{Hicks2019}. 
While audio or haptic modalities have been shown to enhance PX~\cite{juicyhaptic}, visual effects and especially particles remain the primary juicy effect
, used to depict 
elements like fire, explosions, and water~\cite{hastingsParticle1,hastingsParticle2}. However, text-based effects, including onomatopoeia, are an under-explored avenue for improving PX.



Onomatopoeic effects are common in games, like Persona 5 and Valorant, often mimicking comic book or manga aesthetics. However, research on their use beyond comic books and linguistics is scarce. Animated onomatopoeic text effects in videos can clarify sound dynamics, enhance visual impact, and make content more enjoyable to watch~\cite{Wang2017}. In social media videos, they improved content focus~\cite{Zhang2022,Simpson2023}. In games, comic book style onomatopoeia can effectively represent sound effects in subtitles~\cite{SubtitleEmpirical}. The multi-modality of onomatopoeia~\cite{Sasamoto2019Intro} and previous findings on differences between visual and text-based stimuli~\cite{Fabre2021} suggest further research is needed to uncover the full potential of text effects.

Onomatopoeia as a means of using text to express visual pizzazz, i.e., juiciness, raises our first research question: \textbf{RQ1. Do 
juicy text-based effects modeled on onomatopoeia elicit juicy user experiences, similarly to purely visual-based approach?} 
The small collection of findings from comic books, linguistics, and captioning points in this direction, leading us to hypothesize:

\textit{\textbf{H1a.} There will be a significant increase in PX when text-based juicy effects are used vs. when no effect appears.}

and, in recognition of the focus on particle effects, juicy and otherwise, we also hypothesize:

\textit{\textbf{H1b.} There will be a significant increase in PX when text-based juicy effects are used vs. when 
particle effects are used.}

\subsection{Text Juice in Player Performance}

Translating phenomena from one modality to another---even spoken words to a written form, as for onomatopoeia---may induce extra cognitive effort~\cite{guynesOno}, which could then degrade performance. Work on cognitive load theory would suggest that additional text would reduce performance~\cite{jamet2007effect}. In video games, visual clutter, which could include text effect, were shown to have a negative influence on performance~\cite{DELMAS2022103628,Caroux2013}.
Considering this, we ask \textbf{RQ2}. \textbf{Do text-based juicy effects 
degrade user performance?} Since text and especially onomatopoeia comprehension may be slower than for simple shapes and colors, we hypothesize:

\textit{\textbf{H2.} Player performance will significantly degrade for text-based effects (both onomatopoeic and semantic), compared to particle-based effects, which are visual and non-linguistic.}

Text-based juice can also be non-onomatopoeic but still semantic. 
JT content can carry other forms of semantic meaning 
to be parsed by players. A common example in games is "flying text," the small pieces of textual information about user states (damage inflicted, score, ...) that result from their actions (e.g., Diablo, Super Mario Wonder, etc.), though these are not always juiced up. Work on 
film subtitling suggests that meaningful text will not be disruptive and in fact enhance understanding of the content~\cite{perego2010cognitive}. Indeed, \citet{mayer2002multimedia}'s theory of multimedia predicts that complementary text alongside images would be beneficial for understanding.
For VFX in game contexts, 
however, juice may lead players to discount text content in the face of visual zest. We therefore asked: \textbf{RQ3. Does juiciness cause players to overlook the \emph{meaning} of the content represented by text-based juice?}

Juice properties, notably color, can be compelling or distracting~\cite{van1987attraction}. 
Still, the work so far suggests that juicy is uplifting without being distracting~\cite{Hicks2019,juul2016good}. One way we can evaluate this for text content is by providing words with meaning, i.e., \emph{semantic text}, and words without meaning, i.e., \emph{non-semantic or gibberish text}~\cite{smallwood2007counting,schad2012your}. 
Presenting gibberish may, however, cause the player to expend effort on trying to make sense of it~\cite{Sasamoto2016,flyxe2002translation}, i.e., be distracting rather than elicit a juicy effect.
Therefore, we hypothesize:

\textit{\textbf{H3a.} There will be a 
decrease in PX for text without meaning, i.e., gibberish, compared to meaningful text.} 

Putting this in perspective with \textbf{RQ2}, we might also expect degradation of performance with meaningless text compared to meaningful text. We therefore also hypothesize the following:

    \textit{\textbf{H3b.} There will be a decrease in performance for non-semantic text without meaning, compared to semantic text with meaning.}

\subsection{Text Juice and Material Perception}
Another advantage of onomatopoeia may be how it influences the perception of another important visual factor in games: material effects. In the virtual reality (VR) space, \citet{Oh2018} first highlighted the potential of onomatopoeia to enhance UX and immersion. 
The multi-modal capabilities of onomatopoeia 
was found to be able to \textit{alter and add on to the perceived realism/naturalness of the virtual context}~\cite{Choi2018}. However, only non-juiced onomatopoeia were explored. These findings were expended by showing that onomatopoeic text effects can influence the speed of object categorization without affecting the categorization result, as well as alter the perception of object properties~\cite{Fabre2021}. 
In short, juicy onomatopoeia could affect material perception, a key feature of modern video game environments, transmitting information about the world in a clearer manner and therefore impacting PX~\cite{PXI}. Whether these results transfer to non-VR contexts remains unknown. We therefore asked \textbf{RQ4}. \textbf{Do juicy onomatopoeia influence material perception in a game environment, thereby influencing PX?} Based on the non-juicy and non-game findings, we hypothesized:

\textit{\textbf{H4.} People will tend to rate the material properties similarly to the onomatopoeic effect used.}

\section{Research Game for Juicy Text Study}
We discuss the game design and the process of crafting the effects used in our study. We 
focused on visual effects, despite onomatopoeia's association with sound. This allowed a controlled comparison between text and particle effects. 

\subsection{Effect Design} \label{sec:effectdesign}

\begin{figure}[htp]
\centering
\includegraphics[width=.7\columnwidth]{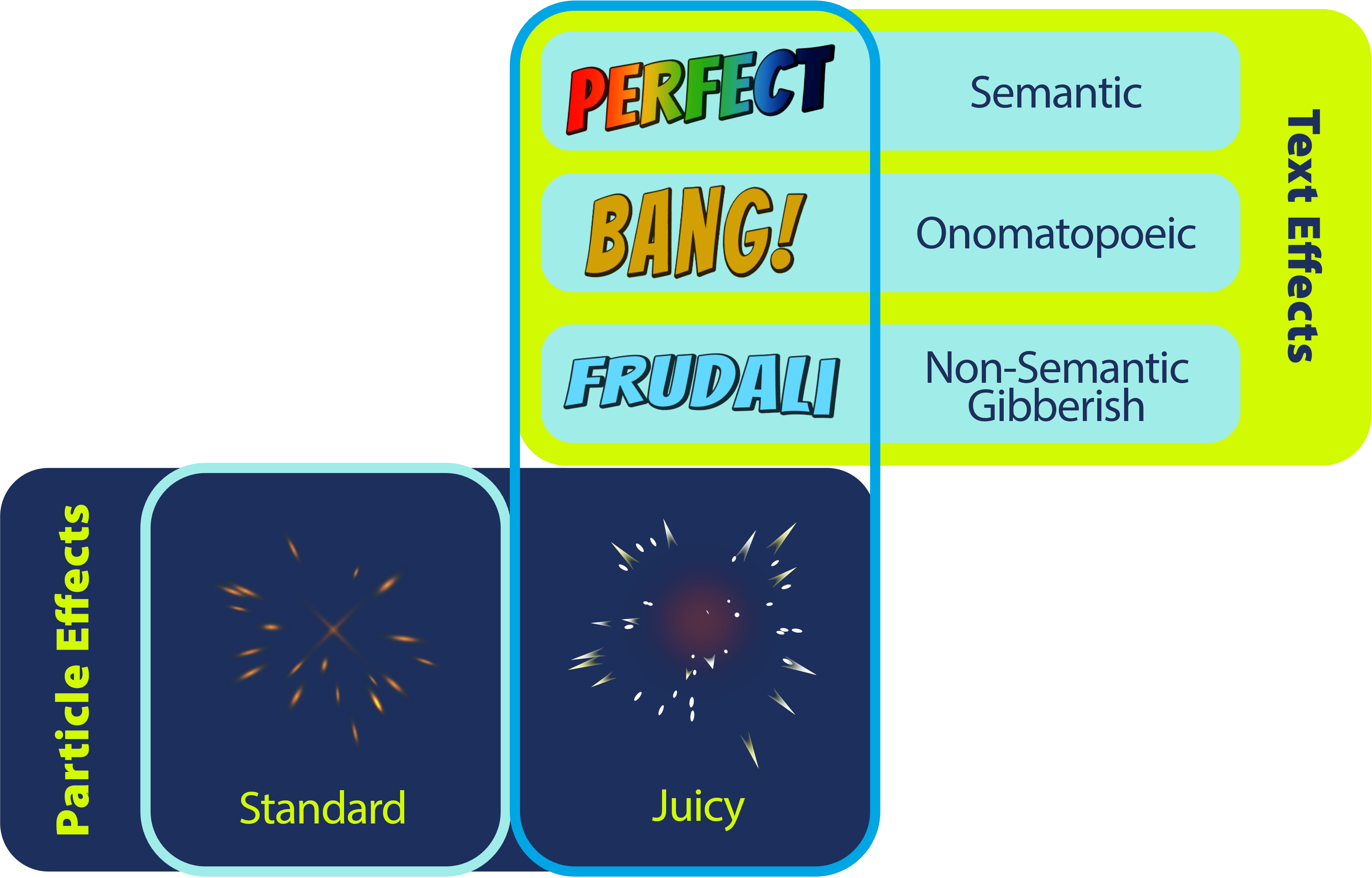}
\caption{Types of effect used in the study.}
\Description{An array of effect separating "Particles" between "Standard" and "Juicy" and "Text" between "Semantic", "Onomatopoeic" and "Non-Semantic/Gibberish". "Text effects" are also on the "Juicy" axis}
\label{fig:effectdiagram}
\end{figure}

We categorized our effects as \emph{Particles} and \emph{Text} (\autoref{fig:effectdiagram}). Text effects were further categorized as \emph{Non-Onomatopoeic Semantic} (hereafter \emph{Semantic}), \emph{Onomatopoeic Semantic} (hereafter \emph{Onomatopoeic}), and \emph{Non-Semantic/Gibberish}. 
All are "Juicy" except for the "Standard" particle effect, 
as per
\citet{Hicks2019}. We mixed linguistic and game design terms. Onomatopoeia are a type of text content. "Juicy" refers to an embellishment effect. "Juicy onomatopoeia" is an embellished text effect where the text is an onomatopoeia.

\begin{figure}[ht]
\captionsetup[subfigure]{justification=centering}
\centering
    \begin{subfigure}[t]{0.175\columnwidth}
        \centering
        \includegraphics[width=\linewidth]{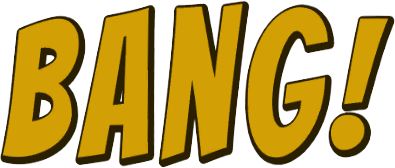}
        \caption{Onomatopoeic explosion.}
        \Description{Yellow "BANG!" text in a cartoonish font.}
        \label{fig:obsstageexamp}
    \end{subfigure}
    \hfill
    \begin{subfigure}[t]{0.175\columnwidth}
        \centering
        \includegraphics[width=\linewidth]{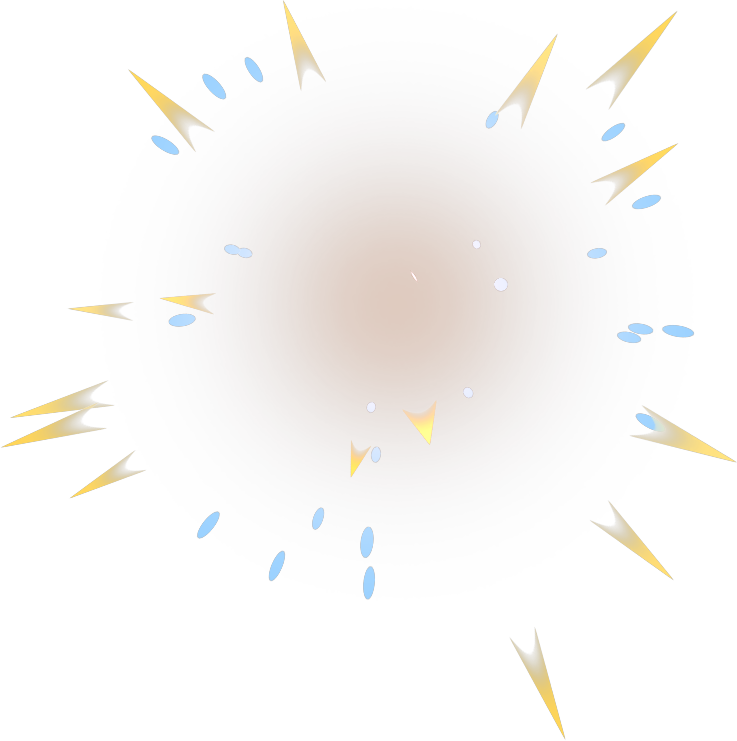}
        \caption{"Juicy" explosion particles.}
        \Description{Orange halo with blue lines and yellow triangle emanating from it.}
        \label{fig:obsstageexamp2}
    \end{subfigure}
    \hfill
    \begin{subfigure}[t]{0.175\columnwidth}
        \centering
        \includegraphics[width=\linewidth]{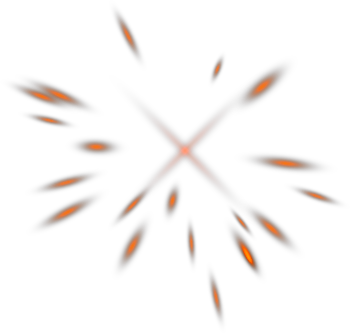}
        \caption{"Standard" explosion particles.}
        \Description{Small orange cross with small orange lines emanating from it.}
        \label{fig:obsstageexamp3}
    \end{subfigure}
    \hfill
    \begin{subfigure}[t]{0.175\columnwidth}
        \centering
        \includegraphics[width=\linewidth]{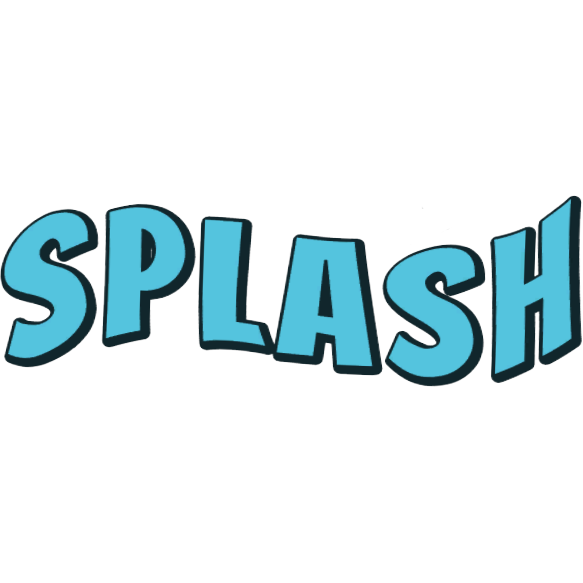}
        \caption{Onomatopoeic water effect.}
        \label{fig:waterono}
        \label{fig:watereff}
        \Description{A blue SPLASH text in a cartoonish font.}
    \end{subfigure}
    \hfill
    \begin{subfigure}[t]{0.175\columnwidth}
        \centering
        \includegraphics[width=\linewidth]{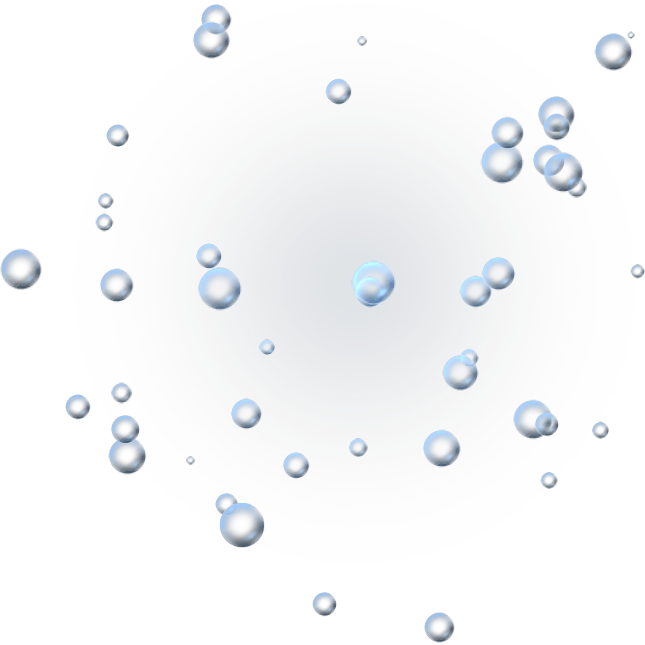}
        \caption{"Juicy" water particle.}
        \label{fig:waterpart}
        \Description{Dark blue halo with bubble like blue circles emanating from it.}
    \end{subfigure}

    
    \hfill
        \begin{subfigure}[t]{0.175\columnwidth}
        \centering
        \includegraphics[width=\linewidth]{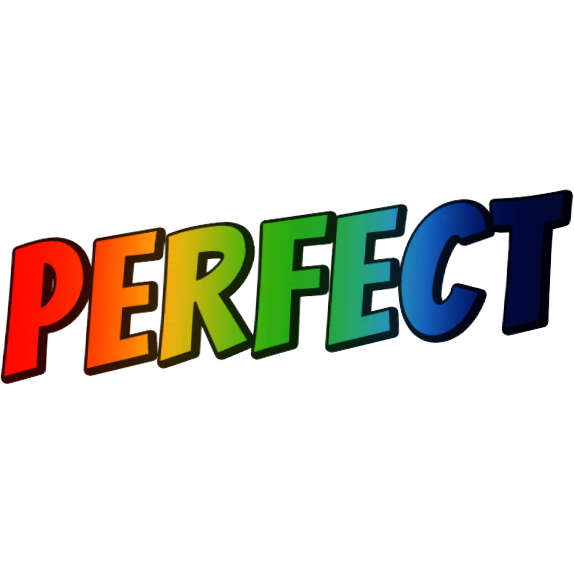}
        \caption{"Perfect" semantic feedback.}
        \Description{A rainbow colored "Perfect" text}
        \label{fig:perfeff}
    \end{subfigure}
    \hfill
    \begin{subfigure}[t]{0.175\columnwidth}
        \centering
        \includegraphics[width=\linewidth]{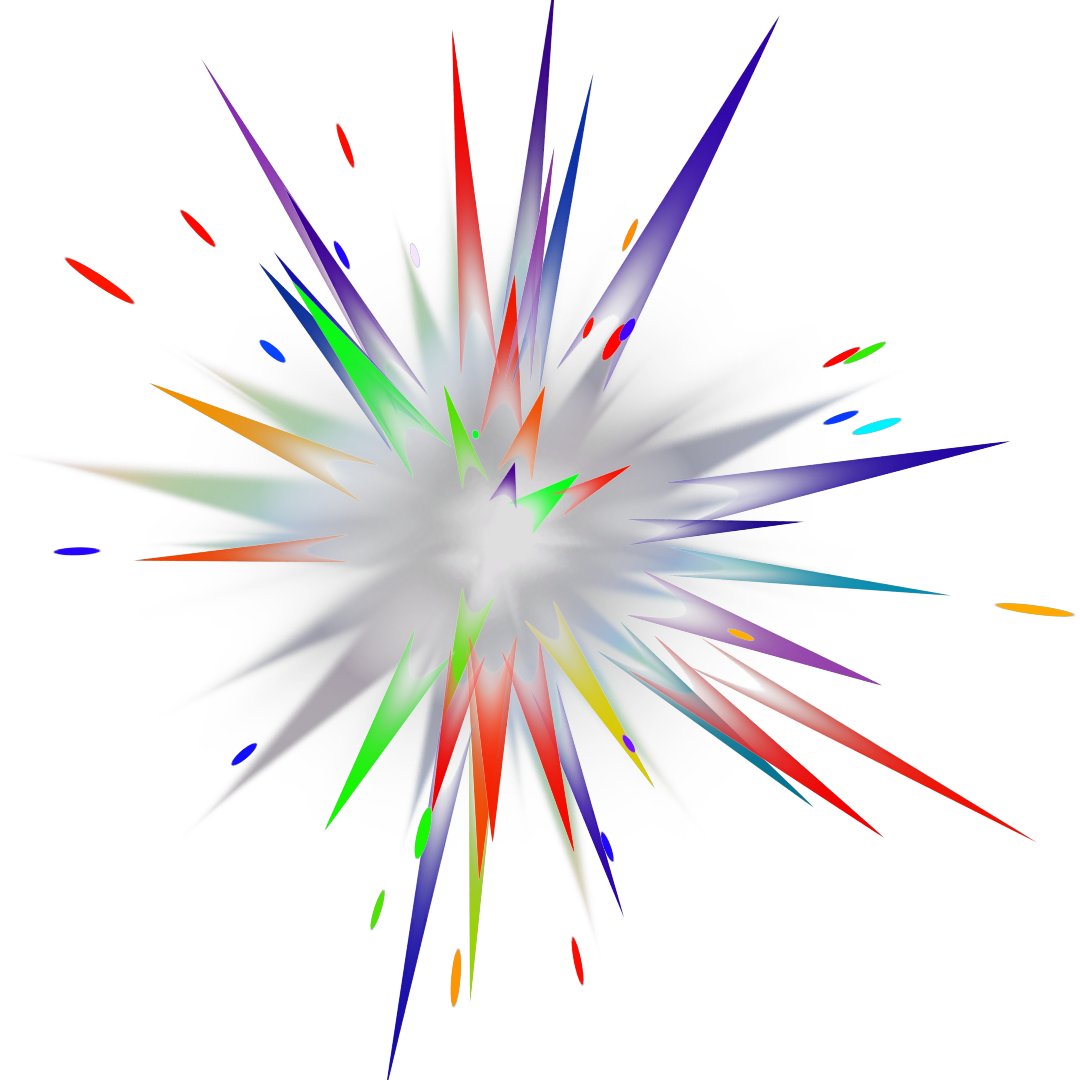}
        \caption{"Perfect" particle feedback.}
        \label{fig:perfex1}
        \Description{A white halo from which rainbow colored triangles and lines are radiating out of}
    \end{subfigure}
    \hfill
    \begin{subfigure}[t]{0.175\columnwidth}
        \centering
        \includegraphics[width=\linewidth]{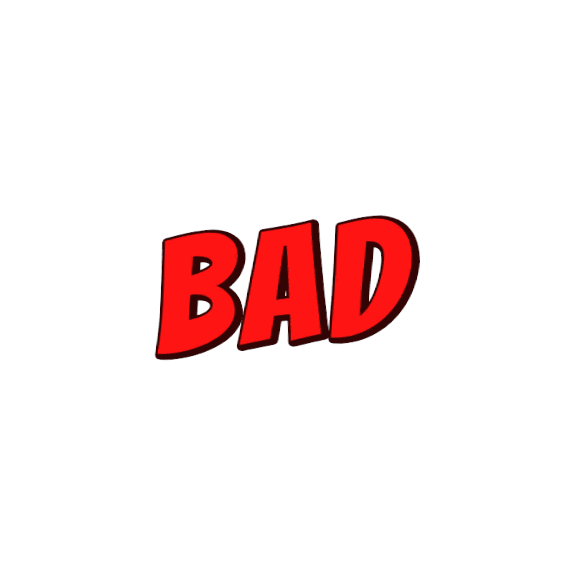}
        \caption{"Bad" semantic feedback.}
        \Description{A red "Bad" text}
    \end{subfigure}
    \hfill
    \begin{subfigure}[t]{0.175\columnwidth}
        \centering
        \includegraphics[width=\linewidth]{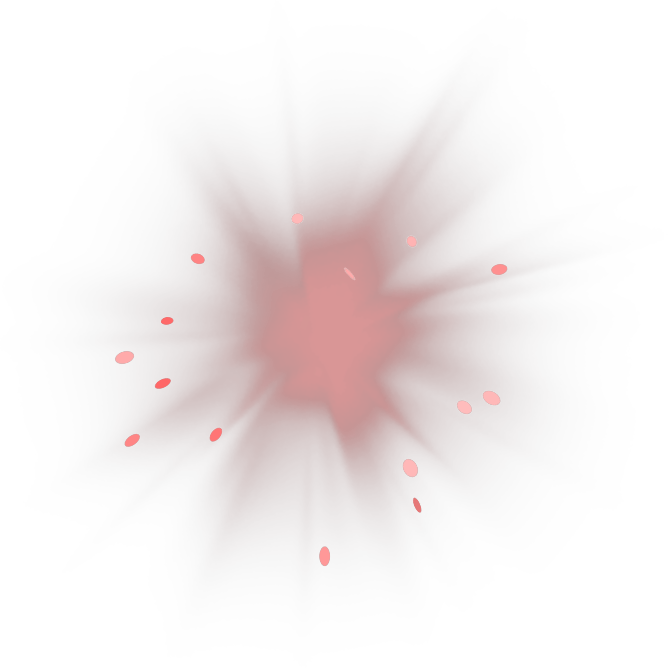}
        \caption{"Bad" particle feedback.}
        \label{fig:perfex2}
        \Description{A red halo with small white triangles radiating out of it}
    \end{subfigure}
 


    \hfill
    \begin{subfigure}[t]{0.175\columnwidth}
        \centering
        \includegraphics[width=\linewidth]{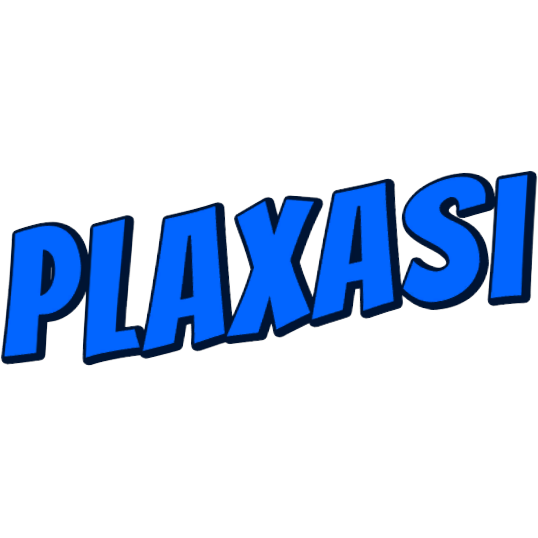}
        \caption{Random text variant A.}
        \Description{Blue "PLAXASI" text in a cartoonish font.}
        \label{fig:matstageexamp}
    \end{subfigure}
    \hfill
    \begin{subfigure}[t]{0.175\columnwidth}
        \centering
        \includegraphics[width=\linewidth]{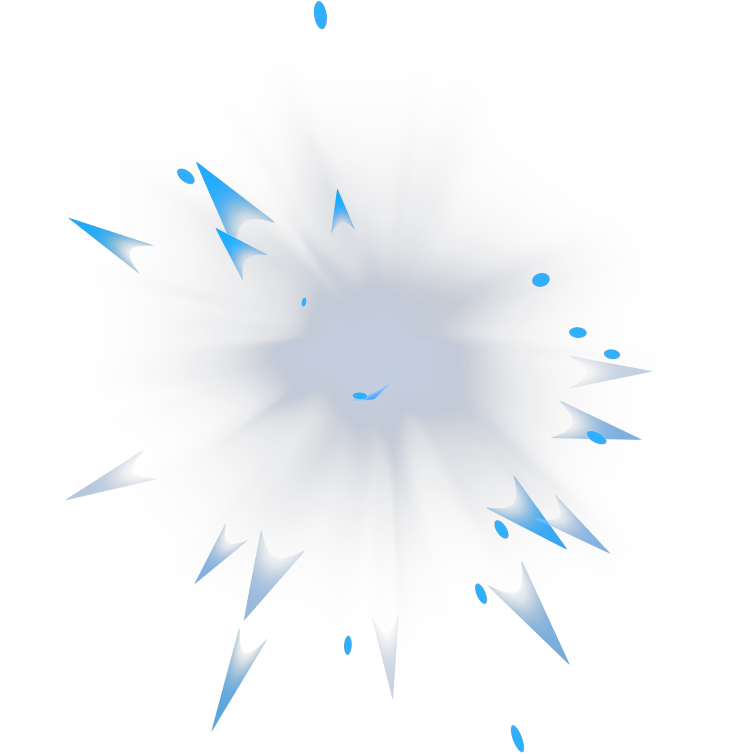}
        \caption{Random particle variant A.}
        \Description{Dark blue halo with blue triangles emanating from it}
    \end{subfigure}
    \hfill
    \begin{subfigure}[t]{0.175\columnwidth}
        \centering
        \includegraphics[width=\linewidth]{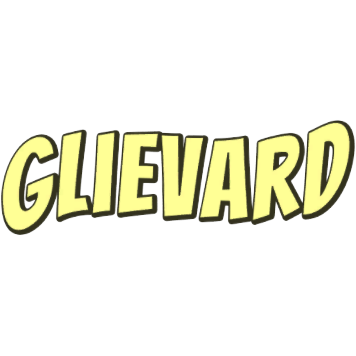}
        \caption{Random text variant B.}
        \Description{Pale yellow "GLIEVARD" text in a cartoonish font}
    \end{subfigure}
    \hfill
    \begin{subfigure}[t]{0.175\columnwidth}
        \centering
        \includegraphics[width=\linewidth]{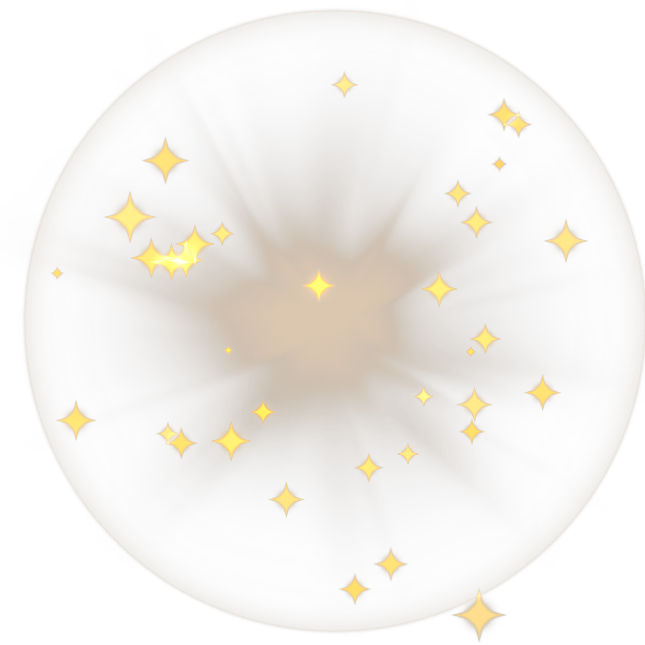}
        \caption{Random particle variant B.}
        \Description{Dark yellow-ish halo with yellow sparkles and a dark yellow-ish circle emanating from it.}
    \end{subfigure}
    \caption{Example of effects used in the study.}   
    \label{fig:alleffects}
\end{figure}

We chose the \textit{Epic Toon FX}\footnote{\url{https://assetstore.unity.com/packages/vfx/particles/epic-toon-fx-57772}} package from Unity's Asset Store as it provided both particle and text-based effects. 
These effects were designed by professionals and highly rated by the Unity community. 
Additionally, for the "Standard" condition, we created basic effects using Unity's VFX Graph\footnote{\url{https://unity.com/visual-effect-graph}}. Each effect lasted half a second, fading in and out smoothly.
The type of effect used most, an explosion, relates to \textit{H1}, \textit{H2}, and \textit{H4}. Effects showcasing "Water" are used to verify \textit{H4}. Non-Semantic/Gibberish Text and randomly selected Particles are used for \textit{H3}.
Semantic Performance feedback are used for \textit{H1a} and \textit{H2}. Refer to \autoref{fig:alleffects} for examples.

\subsection{Minimal FPS Game}
\begin{figure}[htp]
\centering
\includegraphics[width=0.65\columnwidth]{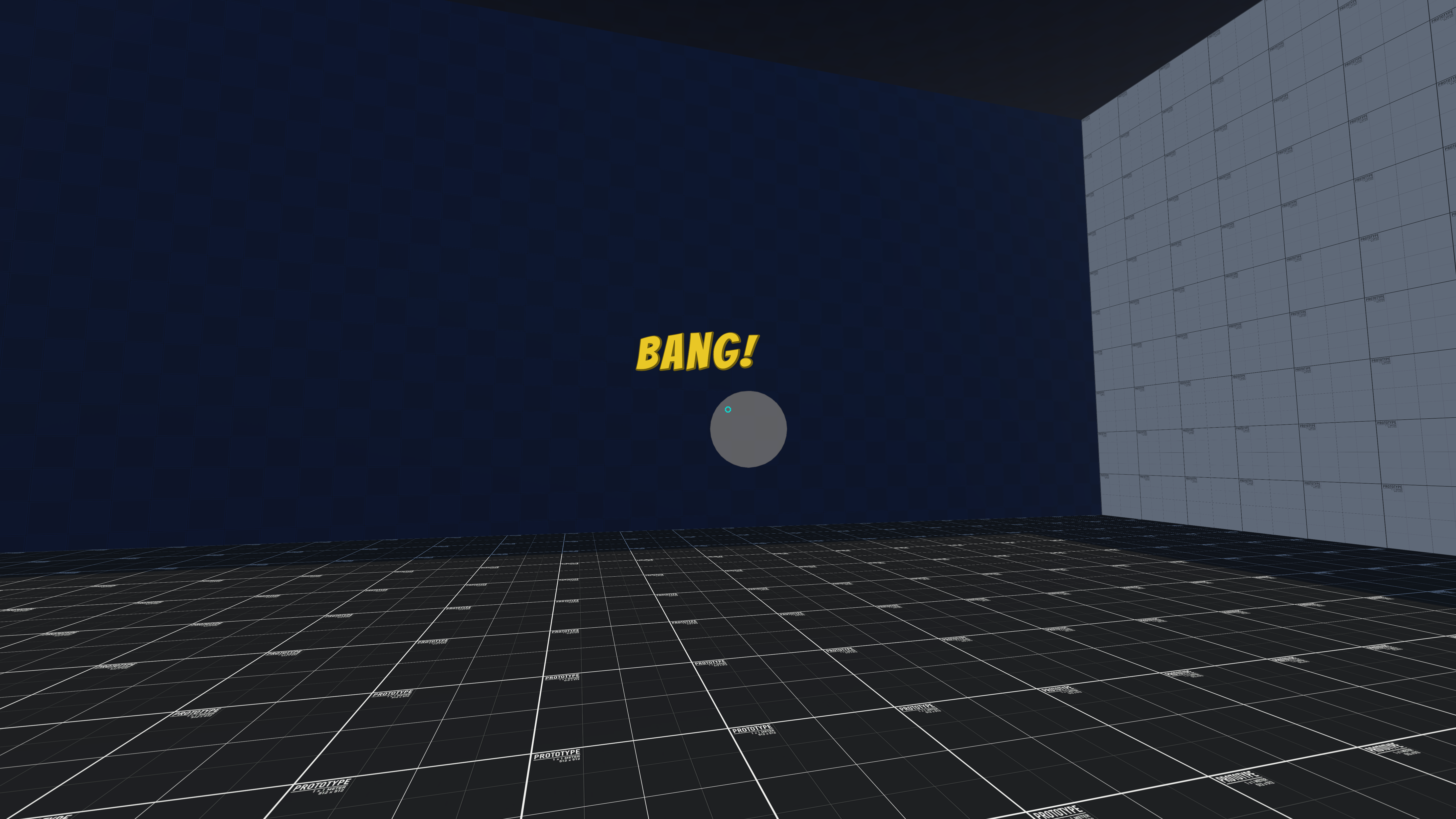}
\caption{Participant view in the experiment.}
\Description{Screenshot from the research game. We see a small room with a roof and a blue wall with a sphere in front of it. A small 2D blue circle is in the middle of the screen to help participant's aim, acting as a crosshair. A yellow BANG! text in a cartoonish font can be seen near the sphere.}
\label{fig:playarea}
\end{figure}

We employed a First Person Shooter (FPS) format, with participants using a standard computer mouse. Known for their clear objectives and structured metrics, FPSs are well-suited for studying the effects of visual effects (VFX) on player experience (PX) and performance \cite{DELMAS2022103628,Caroux2013,Hicks2019}.
The environment (\autoref{fig:playarea}) was an enclosed space with four walls. Players could look around but not move. Spheres randomly spawned near the blue wall, chosen to enhance effect visibility while minimizing distractions~\cite{Caroux2013}. Clicking triggered various effects based on research conditions, such as particle effects or text. 
This "greybox" level-design is a common early-stage development technique in game design \cite{Kramarzewski2023-eq,Chandler2008-dm,hollstrand2020supporting}. Other game-related studies have also utilized this method \cite{STATHAM2022100476,Winters14,juicyhaptic}. 




\section{Overarching Methods}

The study had two phases: (i) an online manipulation check to validate the quality and distinctiveness of the effects (\autoref{sec:ManipCheck}) and (ii) the main experiment (\autoref{sec:MainExp}). The study was approved by the University of Tokyo's ethics board. Analyses were performed using Python's Pingouin package~\cite{pingouin_python}.
First, we cover the online manipulation check study. Then, we report on the experiment, conducted after verifying the manipulation checks.

\section{Manipulation Check Study}
\label{sec:ManipCheck}
We conducted an online manipulation check study to assess the fidelity of effects planned for our experiment \cite{ejelov2020rarely}.

\subsection{Methods}
Participants (N=18) provided keywords and Likert scale ratings for effect videos. Affective reactions were gauged with the Self-Assessment Manikin (SAM) test \cite{SAM}. Keywords like "Liquid" and "Humid" assessed water-like qualities, while others like "Amazing," "Explosive," "Dull," and "Repetitive" validated differences in explosion effects, chosen based on adjectives from an in-lab pilot study. Effects were presented in a random order on a uniform background, named numerically. Participants could replay videos at will. 80\% were aged 18-30, and 20\% were 31-45. All 
received \pounds5. 

\subsection{Results and Discussion}
\label{sec:ManipCheckResults}
The main goal was to confirm our Text effects and the level of juice in our two Particle effects.
We first ran an intraclass correlation (ICC) test to verify consistency in ratings across participants. Results were an average fixed rater ICC (ICC3k) < 0.85 for Repetitive, Humid, and the SAM Valence. 
We therefore chose to not interpret these results. 
The ICC3k of "Amazing," "Explosive," "Dull," and "Liquid" were all > 0.89. 
Data was not normal (Shapiro-Wilk test, p < 0.05). We therefore ran a Friedman test followed by a post-hoc Wilcoxon with Bonferroni corrected p-values.

Water-like effect were statistically significantly perceived as more "liquid", and juicy Particles more "Amazing" and "Explosive", with a higher SAM Arousal, aligning with our expectations. 
Although the lower ratings for Text effects suggests potentially lower appreciation, we proceeded with the main experiment since this check lacked gameplay elements and feedback.

\section{Experimental Design}
\label{sec:MainExp}

We carried out a 
within-participants experiment to compare the impact of type of effect (Particle, Onomatopoeic Text, Semantic Text, ...) and effect element (Explosion, Water, Random) on player engagement, performance, and material perception.

\subsection{Participants}
We used G*Power software~\cite{Faul2009} to determine an adequate sample size for a Repeated Measures ANOVA with an effect size of $f = 0.25$, chosen based on literature indicating small but significant effects of juiciness~\cite{Hicks2019,hicks2019gamification,Kao2020}.
For the more complex Material Stage (with 6 conditions), we find a sample size of $n=44$.

Participants (N = 45) were recruited online primarily from Western, English-speaking countries (AU, CA, NZ, UK, US). Median age: 28 (M = 30.58, SD = 8.42). Gender distribution: 51\% men, 49\% women. Regarding weekly video game usage, 40\% played more than 13 hours, while 21\% played 6 hours or less. We compensated \pounds5.

\subsection{Procedure} \label{sec:procedure}

There were three distinct stages conducted in participants' web browsers, with a median completion time of $\sim$40 minutes, including 2-5 minute breaks between stages. Each stage addressed specific hypotheses and included various conditions with specific effects. Stage order and the order of conditions and questions within stages were randomized. Following each in-stage condition, participants completed an in-game PX questionnaire. The stages were: 

\textbf{Observation Stage (OS):} \label{sec:obsstage-exp} \textit{Conditions: effect Type and Juice.} This stage examined overall player experience across effects, including Standard and Juicy Particle, Juicy Onomatopoeia, Juicy Onoma. + Particle, or No Effect (NE). 
Participants encountered 20 spheres requiring 3 clicks to disappear. Each click spawned an effect.

\textbf{Material Stage (MS):}~\label{sec:matstage-exp} \textit{Conditions: effect Type and Element.} In this stage, the focus shifted to comparing onomatopoeic text effects to particles in conveying information about the world. The conditions swapped the effect's type (Particle or Text) and element like Explosion, Water, and Random (varied shapes and colors with gibberish~\cite{chomsky1956three,gardner1984codes} text). Each condition involved 20 spheres, with participants also responding to various material recognition questions after each conditions.
We generated random words due to the lack of consensus on onomatopoeia lexicability~\cite{flyxe2002translation,Sasamoto2019Intro}. Example includes: "GRAMFID," "POSK," "BLAER." 

\textbf{Performance Stage (PS):}  \label{sec:perfstage-exp} \textit{Condition: effect Type.} Participants 
swiftly clicked spheres with effects based on their timing performance (Slow, Bad, Good, and Perfect, refined through pilot tests). This stage primarily examined non-onomatopoeic semantic text effects for feedback, alongside Particles or NE. Sphere size auto-adjusted based on performance after the 20\textsuperscript{th} sphere to prevent indefinite continuation. Faster participants had shrinking spheres
with "Slow" and "Bad" feedback, while slower ones had larger spheres.

\subsection{Instruments and Measures}
\label{sec:instrumentsmeasures}\label{sec:pxipres}\label{sec:matpres}
We used a combination of self reports via questionnaires delivered during the game experience and system metrics that captured participant performance to evaluate our dependent variables.

\textbf{Player Experience Questionnaire.} 
A common approach when studying juiciness is to use a PX measure~\cite{hicks2019gamification,Kao2020,juicyhaptic,juul2016good}. We chose to use the Player Experience Inventory (PXI)~\cite{PXI}, a recent and validated measure already used previously by research on juicy effects~\cite{Hicks2019,juicyhaptic}. Each of the 10 subscale of the PXI correspond to an aspect of gameplay (i.e. Audiovisual appeal, Clarity of goals, ...). 
Participants used the standard -3 to 3 PXI response scale during the experiment, but we later use the standard form for Likert scales in the results, i.e., a 1 to 7 scale.

\begin{figure}[!ht]
\centering
\includegraphics[width=0.8\linewidth]{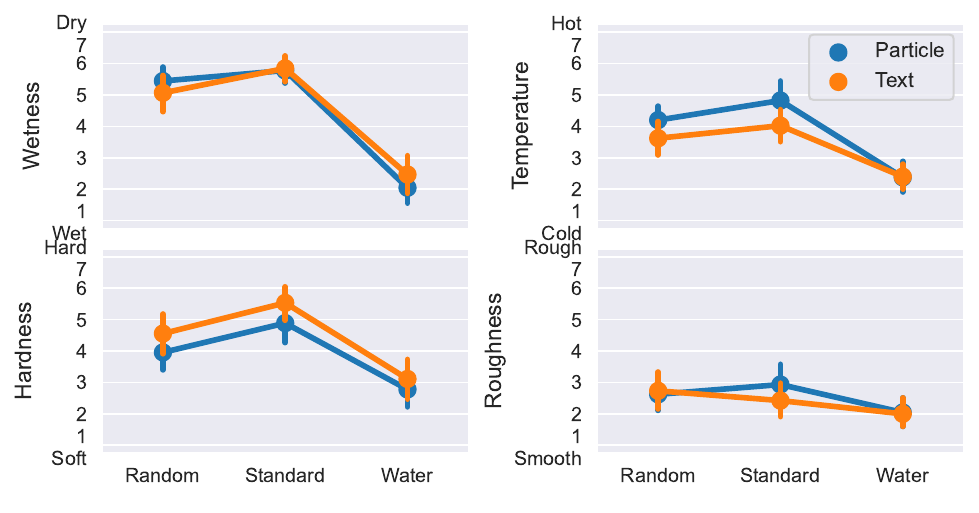}
\caption{Material evaluation results.}
\label{sec:matgraph}
\Description{4 point plots shows participant evaluation of the sphere based on the effect and on the conditions. 2 sets of points exist, one for Particle effects one for Text effects.
Wetness: Ordinate: 1 (Wet) to 7 (Dry). Abscissa: Effect material, from left to right: Random, Explosion, Water. Point for Random and Standard hover around 5 while they are closer to 3 for the Water text and 2 for the Water particle.
Hardness: Ordinate: 1 (Soft) to 7 (Hard). Abscissa: Effect material, from left to right: Random, Explosion, Water. Almost no difference between Text and Particle. Random is around 4. Explosion is around 5.5. Water is around 3.
Roughness: Ordinate: 1 (Smooth) to 7 (Soft). Abscissa: Effect material, from left to right: Random, Explosion, Water. Text is always slighly higher than particles. Random: Around 4. Explosion: Close to 4 for particles, close to 5 for text. Water: Around 3 for text, around 2.5 for particles.
Temperature: Ordinate: 1 (Cold) to 7 (Hot). Abscissa: Effect material, from left to right: Random, Explosion, Water. Random: Text around 4, Particles slightly lower than 5. Explosion: Particle is around 5.5, text is around 4. Water: Both particles and text are around 2.}
\end{figure}

\textbf{Material Recognition.} 
In the MS, we base our material recognition questionnaire on the framework used by \citet{Fabre2021}. We asked to rate the sphere Wetness, Hardness, Roughness, and Temperature on a 1 to 7 Likert scale, with the 1 and 7 value corresponding those presented in \autoref{sec:matgraph}.

\textbf{Performance Metric.}
We focused on speed and accuracy. 
Time-to-click measured the duration to hit each sphere individually, while time between multi-click measured the interval between clicks for spheres requiring it. 
Accuracy is linear from 0 to 1, with 1 indicating no misses.
These metrics, akin to reaction time and accuracy, are commonly used in performance related research~\cite{DELMAS2022103628,Caroux2013,beck2012searching,Azizi2016}.

\subsection{Data Analysis}
For each stage, we took the overall score of each PXI construct per participant. Each construct was analyzed independently.
For performance-related data on the PS, we only took into account the first 20 spheres, as the size nudge started to take effect from the 21\textsuperscript{st} sphere onwards. 
We ran Shapiro-Wilk tests of normality on the data. We also considered QQ plots. We used repeated measures ANOVAs followed by pairwise t-tests with Bonferroni corrections for normal data and Friedman tests followed by Wilcoxon signed-ranked tests with Bonferroni corrections 
for data that violated normality. We used $\alpha < 0.05$ for significance and *$p < .05$, **$p < .01$, ***$p < .001$.


\section{Results}


\begin{figure*}[!ht]
\centering
  \includegraphics[width=.85\linewidth]{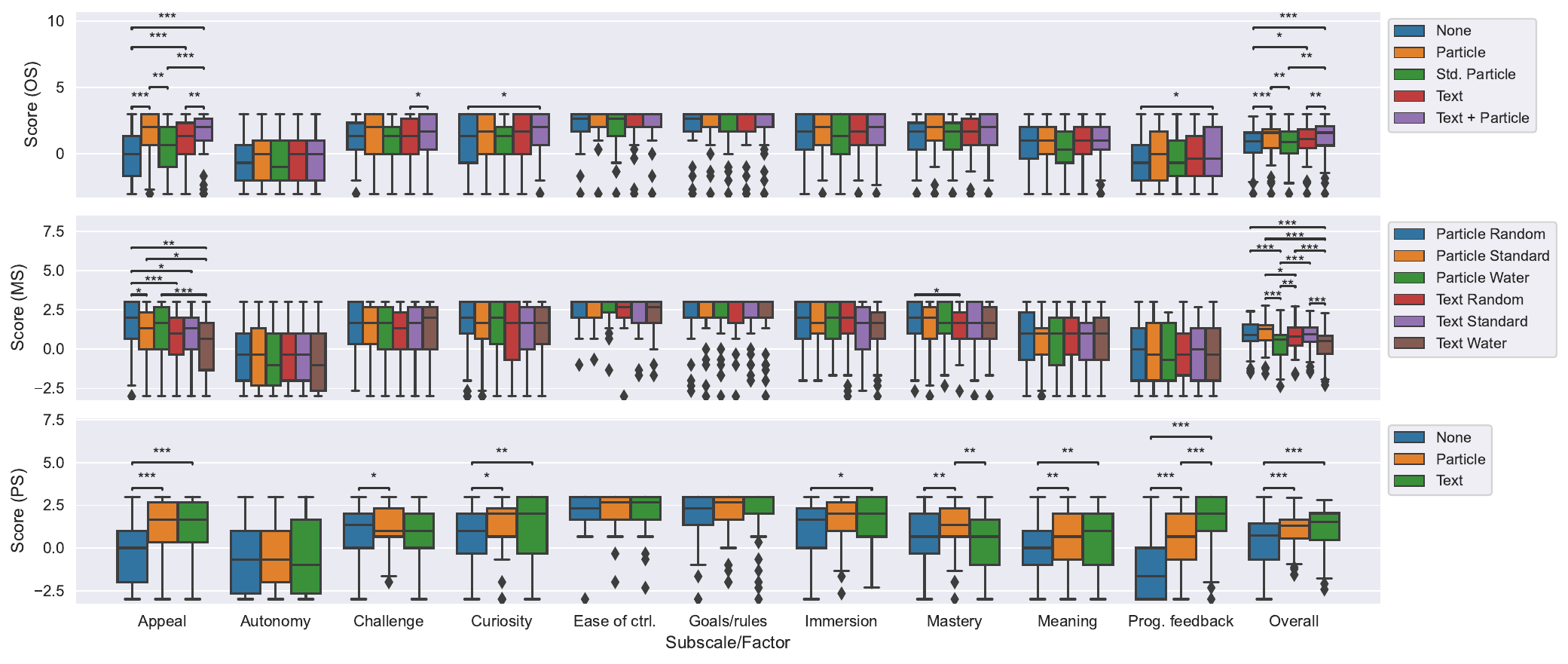}
  \captionsetup{justification=centering} 
\caption{PXI results for all stages. OS: Observation Stage. MS: Material Stage. PS: Performance Stage.}
  \label{fig:pxi_graph}
  \Description{The image is a series of boxplots, each representing a different construct from the Player Experience Inventory (PXI). The boxplots are color-coded to represent different types of responses: blue for "None," orange for "Particle," green for "Std. Particle," and purple for "Text + Particle.". Different colors are used for the Material Stage and Performance Stage due to the different conditions. Each boxplot shows the distribution of scores for a particular construct, with the central box representing the middle 50\% of the data, the whiskers showing the range of the data, and the dots representing outliers. The constructs are labeled as "Appeal," "Autonomy," "Challenge," "Curiosity," "Ease of control," "Goals and rules," "Immersion," "Mastery," "Meaning," and "Progress feedback." The boxplots are arranged in a grid format, with each row representing a different construct and each column representing a different stage of the experiment. The median scores and the spread of the data can be compared across the different constructs and color categories. Results are similar between Observation Stage and Material Stage, but change for the Performance Stage with text being usually higher. Ease of ctrl and Goals and rules are the highest rated construct with the lowest variation accross all stages and conditions, with Appeal, Autonomy, Challenge, Curiosity and Meaning spanning accross the whole -3/3 range.}
\end{figure*}

\subsection{Text and Juice Impact on PXI (H1a, H1b)}
\label{sec:textpxires}

\subsubsection{Observation Stage}
\begin{table}[ht]
    \centering
    \caption{Summary of PXI results for the Observation Stage. N: No Effect. T: Text. P: Juicy Particles. \=P: Standard Particles. We only show statistically significant results.}
    \resizebox{0.9\columnwidth}{!}{%
    \begin{tabular}{llll}
    \toprule
        \textbf{Subscale} &\textbf{Friedman} & \textbf{Wilcoxon} \\
    \midrule
       Overall & W = .21, $\chi^2$(4) = 38.44, p < .001 *** & \multirow{2}{*}{\shortstack[l]{P, T, P+T > N,\\P > \=P; T+P > \=P, T}}\\
       &&&\\
        Appeal & W = .26, $\chi^2$(4) = 46.49, p < .001 *** & \multirow{2}{*}{\shortstack[l]{T+P > T > N,\\T+P > \=P; P > \=P, N}} \\
       &&&\\
        Challenge & W = .10, $\chi^2$(4) = 17.47, p = .001 ** & T+P > T \\
        Curiosity & W = .07, $\chi^2$(4) = 12.33, p = .01 * & T+P > N \\
        Prog. feedback & W = .06, $\chi^2$(4) = 10.87, p = .02 * & T+P > N \\
    \bottomrule
    \end{tabular}
    }
    \label{tab:step1res}
\end{table}

Statistically significant effects were found, with Appeal and Challenge showing small differences (Kendall’s W > 0.1). (refer to \autoref{tab:step1res}). 

    \textbf{Appeal construct.} We found that the \textit{Text+Particle} (M = 5.51, SD = 1.81) were rated higher than all other effects, except \textit{Juicy Particles} (M = 5.34, SD = 1.75). \textit{Juicy Particles} were rated higher than \textit{Standard Particles} and \textit{NE}. \textit{Text} was rated higher only to \textit{NE}. 
    
    \textbf{Challenge construct.} \textit{Text+Particle} (M = 5.28, SD = 1.74) was rated higher than \textit{Text} (M = 5.01, SD = 1.80), with $W = 60.5$, $p = .03$.
    
    \textbf{Curiosity and Progress feedback.}
    Results also showed \textit{Juicy Particle} and \textit{Text+Particle} effects being statistically significantly better than \textit{NE}, with $W = 31$ and $161.5$, and $p = .03$ and $.04$ for Curiosity and Progress feedback, respectively.

\subsubsection{Performance Stage}

\begin{table}[ht]
    \centering
    \caption{PXI results for the Performance Stage. N: No Effect, T: Text, P: Particles. We show statistically significant results.}
    \resizebox{0.85\columnwidth}{!}{%
    \begin{tabular}{llll}
    \toprule
        \textbf{Subscale} &\textbf{Friedman} & \textbf{Wilcoxon} \\
    \midrule
       Overall & W = .18, $\chi^2$(2) = 16.13, p < .001 *** & P > N; T > N\\
        Appeal & W = .27, $\chi^2$(3) = 24.78, p < .001 *** & P > N; T > N \\
        Challenge & W = .07, $\chi^2$(2) = \phantom{0}6.83, p =  .03 * & P > N \\
        Curiosity & W = .09, $\chi^2$(2) = \phantom{0}8.12, p = .01 * & P > N; T > N \\
        Immersion & W = .11, $\chi^2$(2) = 10.17, p = .006 ** & T > N \\
        Mastery & W = .18, $\chi^2$(2) = 16.61, p < .001 *** & P > T, N \\
        Meaning &W = .23, $\chi^2$(2) = 20.48, p < .001 ***  & P > N; T > N \\
        Prog. feedback & W = .44, $\chi^2$(2) = 39.91, p < .001 *** & T > P > N \\
    \bottomrule
    \end{tabular}
    }
    \label{tab:step3res}
\end{table}

Friedman tests were statistically significant across all constructs except Autonomy, Ease of Control, and Goals \& Rules (refer to \autoref{tab:step3res}).
While the Appeal construct showed a similar result to the OS, other performance- and mastery-focused constructs showed a larger effect. 
By construct:

    \textbf{Appeal construct.} \textit{Particles} (M = 5.21, SD = 1.79) and \textit{Text} (M = 4.98, SD = 1.91) were higher than \textit{NE} (M = 3.76, SD = 1.87), $W = 71.5, p < .001$ and $W = 80, p < .001$ for P>N and T>N respectively.
    
    \textbf{Challenge.} \textit{Particles} (M = 5.26, SD = 1.31) were rated higher than \textit{NE} (M = 5.26, SD = 1.31), $W = 139, p = .01$
    
    \textbf{Mastery construct.} We found \textit{Particle} (M = 5.24, SD = 1.48) higher than both \textit{Text} (M = 4.38, SD = 1.84) and \textit{NE} (M = 4.65, 1.74), $W = 123.5, p = .003$ for P>N and $W = 110, p = .001$ for T>P. 
    
    \textbf{Curiosity and Meaning constructs.} Both \textit{Particle} (Cur.: M = 5.28, SD = 1.75; Mea.: M = 4.46, SD = 1.9) and \textit{Text} (Cur.: M = 5.25, SD = 1.97; Mea.: M = 4.56, SD = 1.94) improved results vs. \textit{NE} (Cur.: M = 4.55, SD = 1.96; Mea.: M = 3.95, SD = 1.73). Cur.: $W = 75.5, p = .004$ and $W = 128, p = .02$; Mea.: $W = 131.5, p = .008$ and $W = 111, p = .004$ for P>N and T>N, respectively.
    
    \textbf{Immersion.} \textit{Text} (M = 5.47, SD = 1.64) was higher than \textit{NE} (M = 4.95, SD = 1.88), $W = 143.5, p = .02$. 
    
    \textbf{Progress feedback.} We found difference between all pairs, with \textit{Text} (M = 5.62, SD = 1.56) rated higher than \textit{Particle} (M = 4.43, SD = 1.97), and both are higher than \textit{NE} (M = 2.81, SD = 1.85), $W = 101$ for P>N, $22$ for T>N, and $105$ for T>P.

We \textbf{\textit{accept \textit{H1a} but reject \textit{H1b}}}, as \textit{Text} failed to improve PX on its own compared to \textit{Particles}, but still fared better than \textit{NE}.

\subsection{Text Effect Content Impact on the PXI (H3a)}
\label{sec:textcontentresult}

Analysises were conducted to compare effect type and element of the MS. PXI results are similar to the OS 
(refer to \autoref{tab:step2res}).

\begin{table}[ht]
    \centering
    \caption{Statistically significant results for the Material Stage. P: Particle, T: Text, S: Standard.
    W: Water. R: Random. ">" = "more" (i.e., PW > PR means PW was more damp than PR). 
    }
    \resizebox{0.85\columnwidth}{!}{%
    \begin{tabular}{llll}
    \toprule
        \textbf{Subscale} &\textbf{Friedman} & \textbf{Wilcoxon} \\
    \midrule
       Overall & W = .41, $\chi^2$(5) = 93.58, p < .001 *** & \multirow{2}{*}{\shortstack[l]{S, R > W\\PS > TR}} \\
       &&&\\
       Appeal & W = .18, $\chi^2$(5) = 37.71, p < .001 *** & \multirow{2}{*}{\shortstack[l]{PR > S, TR, TW\\PS, PW > TW}} \\
       &&&\\
       Mastery & W = .09, $\chi^2$(2) = 19.83, p = .001 ** & PR > TR \\
    \midrule
    Dampness & W = .49, $\chi^2$(2) = 111.55, p < .001 *** & W > S, R\\
       Hardness & W = .24, $\chi^2$(5) = 55.32, p < .001 *** & \multirow{2}{*}{\shortstack[l]{S > PR; PR > PW\\PS, TS, TR > W}} \\
       &&& \\
       Temperature & W = .25, $\chi^2$(2) = 56.83, p <  .001 *** & S, R > W; PS > TR\\
    \bottomrule
    \end{tabular}
    }
    \label{tab:step2res}
\end{table}

\textbf{Appeal construct.} \textit{Random Particles} (M = 5.78, SD = 1.38) are statistically significantly favored vs. every other effect except \textit{Water Particles} (M = 5.6, SD = 1.41).  We also found that \textit{Standard Particles} (M = 5.52, SD = 1.58) and \textit{Water Particles} were rated higher than \textit{Water Text} (M = 5.38, SD = 1.49). 

\textbf{Mastery construct.} \textit{Random Particles} (M = 5.78, SD = 1.38) rated higher than \textit{Random Text} (M = 5.39, SD = 1.26). $W = 85, p = .01.$

We must \textbf{\textit{reject hypothesis \textit{H3a}}}, since we observed no significant difference between Gibberish Text and other types of text. Gibberish may not always be ideal in terms of design, but the use of such Non-Semantic Text did not degrade PX.

\subsection{Text Effect and Material Perception (H4)}
\label{sec:matres}

Our material perception measures lacked validation despite being based on previous research~\cite{Fabre2021}. We therefore used ICC to assess consistency between users. Results showed high consistency (ICC3k > 0.9) for all keywords except Roughness. Friedman tests revealed differences across Wetness, Hardness, and Temperature, with Dampness showing the largest effect (W = 0.49). Post-hoc tests (\autoref{tab:step2res}) found both Water Particle and Text effects significantly closer to descriptors like "Wet" and "Cold." However, no difference was found between the two. 
This indicates that changing the text element significantly affects sphere perception, \textbf{confirming \textit{H4}}.

\subsection{Text Effect and Performance (H2, H3b)}
\label{sec:playerperf}


For the OS and PS, Friedman tests showed no statistically significant differences between effects. However, the MS shows a difference for Accuracy and Time between clicks. Post-hoc tests show that \textit{Explosion Text} (M = 0.9, SD = 0.1) yielded significantly better accuracy than \textit{Water Particles} (M = 0.85, SD = 0.13), $p = .03$. Additionally, non-random \textit{Particles} (both M = 0.17, SD = 0.03) resulted in significantly shorter time between clicks vs. \textit{Random Text} (M = 0.2, SD = 0.05), $p = .04$.

Despite expecting text to reduce performance compared to particles, there was no significant impact, leading us to \textbf{reject \textit{H2}} for both Semantic and Onomatopoeic Text. However, Gibberish did impact performance, though to a lesser extent than anticipated, as evidenced by the significant difference between \textit{Random Text} and \textit{Particles} in time between clicks. Thus, we partially \textbf{accept \textit{H3b}}.

\begin{figure}[ht!]
    \centering
  \includegraphics[width=.8\columnwidth]{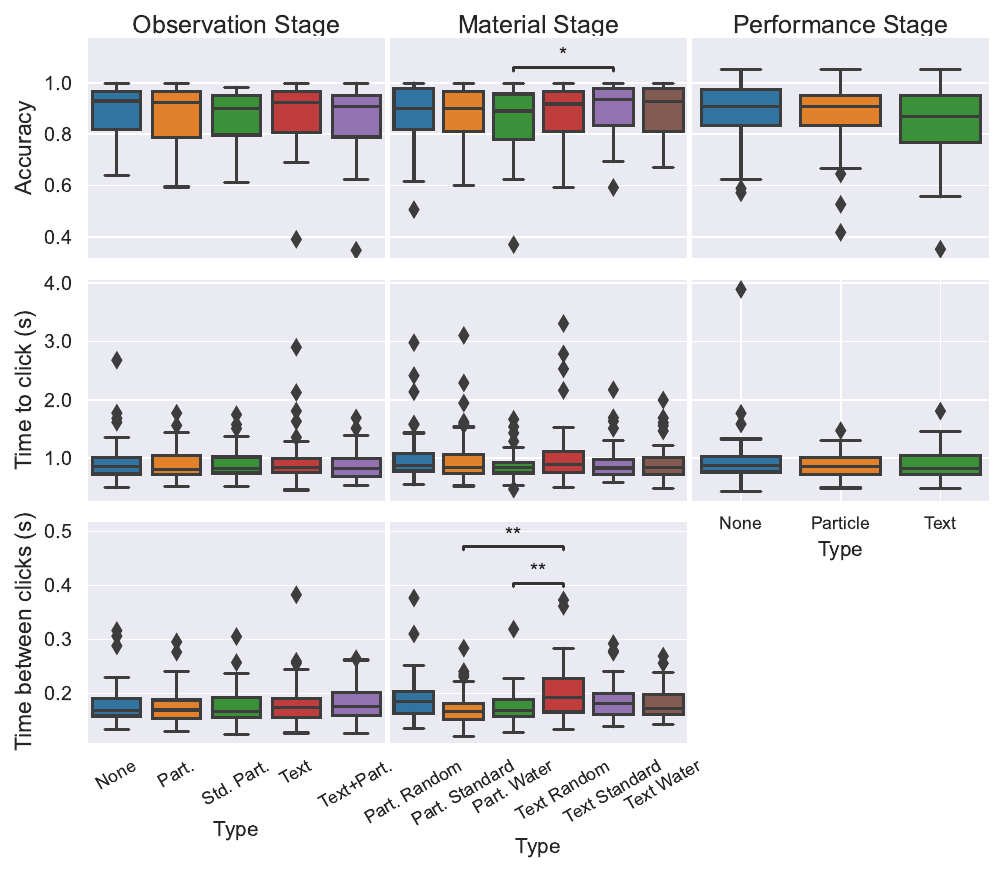}
  \captionsetup{justification=centering} 
\caption{Performance results for all stages. s: Seconds.}
\Description{This image is a series of boxplot representing the performance across all of the experiment's stages. We see little difference on the graph themselves, all have close median values but varying degrees of variation with text usually varying the most.}
  \label{fig:perf_graph}
\end{figure}

\subsection{User Preference}
\label{sec:userpref}

After the experiment, we asked participants their preferred effect: 44.4\% chose \textit{Particle}, 40\% \textit{Text+Particles}, 11.1\% \textit{Text}, and 4.4\% \textit{NE}. In a similar question, 48.9\% would choose \textit{Text+Particles}, 33.3\% \textit{Particles}, 13.3\% \textit{Text}, and 4.4\% \textit{NE} if they had to replay with only one effect. Regarding the PS, 62.2\% preferred \textit{Particle}, 31.1\% \textit{Text}, 2.2\% \textit{NE}, and 4.4\% expressed a preference for \textit{Text+Particles}, with answers like "both particle and text" or "
particle and text every time."

\section{Discussion}

Juiciness, including \textit{JT}, partially supported PX enhancement in Appeal. However, \textit{JT} alone only marginally improved PX vs. \textit{No Effects}. Notably, Semantic Text provided more informative feedback in the PS. ALso, combining \textit{JT} 
with \textit{Particles} did not hinder performance and led to improved PX. These findings suggest future research and applications for juiciness and text.

\subsection{The Effects of Juicy Text on PX (RQ1)}

\subsubsection{Onomatopoeic Text}
Aside from audiovisual appeal, we found no significant improvements with juicy onomatopoeic text effects in terms of PX in OS. Similar findings were reported by other studies who also employed a minimalist game approach~\cite{juicyhaptic}.
In contrast to \citet{Hicks2019}, we observed no significant difference between our \textit{Standard} and \textit{Juicy} effects outside of Appeal, while they did find significant differences in Curiosity, Meaning, and Immersion. The base game may play a role~\cite{Kao2020}, with ours being more performance-focused and minimalist rather than exploratory and meaningful. Therefore, JT's impact may depend on how it aligns with the game context, as evident in our PS (\autoref{subsubsec:semantic}), where fitting feedback as juicy \textit{Semantic Text} gave better outcomes.

Ultimately, we cannot recommend using purely visual juiced-up Onomatopoeic Text effects without particle. \textit{Text}, particularly in the Appeal construct, only outperformed \textit{NE}, whereas \textit{Juicy Particles} were at least rated higher than their non-juicy counterparts.

\subsubsection{Semantic Text}
\label{subsubsec:semantic}
In the PS, Juicy \textbf{\textit{\textit{Semantic Text} effects performed as well as juicy \textit{Particles} in most factors}} vs. \textit{NE}, which was not the case in the OS. This difference may be explained by the focus on performance and the better fit of these effects to the overall game design~\cite{Kao2020}. 
The higher results for Progress Feedback may relate to the results of the Mastery construct, with \textit{JT} being lower than both \textit{Particles} and \textit{NE}. This is likely due to the player's actual performance being reported more clearly, such as "BAD" or "SLOW" on the negative side. Perceptions of mastery understandably decreased due to a stronger display of negative feedback.
However, 
the difference between the two may be due to preference or stylistic choice (refer to~\ref{sec:userpref}). 

\subsubsection{Juicy Text}
Contrary to particles, which may be a one-size-fits-all solution, JT 
may need to be more carefully designed to reach its full potential. Still, in the OS, \textbf{\textit{Text did not push Particle effects over the line of "too much" juice,}} as found in \citet{Kao2020}. No statistically significant differences were observed between the juicy \textit{Particles} and juicy \textit{Text + Particles} conditions. The two can coexist and may even benefit from the presence of the other: particles boosting the text's impact and its content potentially influencing the particle effect, although the latter is left to future work.

\subsection{Text Impact on Player Performance (RQ2)}
While we expected that the additional cognitive workload of text processing~\cite{jamet2007effect} would reduce player performance, 
results in \autoref{sec:playerperf} indicate that this was not the case.
This aligns with \citet{Hicks2019}, who also did not find any significant results when adding particle juiciness.
While this does not mean that text and particle effects go through a similar perception process, it does suggest that the difference between the two is insignificant, at least in our case. We want to highlight the complexity of the relationship between effects, PX, and performance~\cite{DELMAS2022103628}. Further research is needed to gain a deeper understanding of these dynamics.

\subsection{Importance of Text Content (RQ3)} 
Comparing \textit{Random Juicy Particle} to \textit{Random Juicy Text}, we observed a significant difference: \textit{Random Particles} boosted the Appeal construct of PXI vs. \textit{standard} ones, 
but \textit{Non-Semantic Text}/Gibberish and \textit{Random Text} did not. 
This could be due to extraneous material having little influence on visual perception~\cite{gilbert1959}. Participant reading speed, potentially affected by repeated exposure to effects, might explain this phenomenon. Text effect \textbf{\textit{content}} seemed to have limited impact on both PX and performance, although \textit{Juicy Random Text} notably increased the time between clicks on the sphere, possibly due to users taking more time to process unusual stimuli. 

While \textit{Random Text} didn't show significant PX differences, juicy \textit{Random Particle} effects notably enhanced PX in the MS, possibly due to their more colorful and dynamic nature. Randomness is recognized as an engaging design feature~\cite{leong2006randomness,liang2012designing,dembski1991randomness} and an essential component of game design~\cite{zhang2021effect}. While it is frequently employed to procedurally generate content~\cite{smelik2009survey}, little is known on the VFX side. 
Despite similar juicy animations and colors, \textit{Random Text} had lower Mastery, Appeal and Time between click than \textit{Random Particles}, further suggesting the significance of text content.


\subsection{Text Impact on Material Recognition (RQ4)}

Both \textit{Particle} and \textit{Text Effects} were effective in conveying the intended perceptions of wetness, temperature, and hardness. 
While is it hard to draw conclusions for the impact on PX, these results hint that onomatopoeic text effects alone \textit{do} influence material perception.
These results reproduce those a VR experiment\cite{Fabre2021} in a non-VR context, while also suggesting that the ``visual descriptors'' (i.e., Particles) they used could have elicited the same results without text. However, this needs to also be explored in a VR environment.

\subsection{Implications for Practice}

We offer some initial implications for game design research and practice based on our findings. While some results aligned with previous work, others did not. We suggest how to apply these insights and prompt future work on the effectiveness of JT.

 \textbf{
    Default to Juicy Particle Effects}: Juicy Particle effects were the preferred option for conveying information.
    While JT can provide better feedback, text alone offers few advantages. Particles can thus be the default option, ideal for most players. 

\textbf{Combine Text with Juiced-up Particles}: If text is used, best combine it with particle effects for appeal.

\textbf{The Advantages of Juicy Text}: JT effects, when combined with Particles, seem to boost player Curiosity, felt Challenge, and especially the usefulness of the feedback received through the text content. 
    JT effects can be more than a callback to comic book aesthetics, despite how they are commonly employed in the industry. 

\textbf{Visual Appeal is Key}: Appeal was the most single impacted subscale in this study.
    We therefore recommend the AttrakDiff~\cite{attrakdiff} to discriminate between similar effects, as it can also help distinguish between Juicy and Standard effects~\cite{Hicks2019}. 

\textbf{Low Impact on Objective Performance}: When keeping to meaningful text, juiced-up text effects do not seem to influence player's objective performance. This falls in line with previous work on juiciness \cite{Hicks2019, hicks2019gamification}. However, future works needs to explore these results, especially with expert players~\cite{DELMAS2022103628}.

\textbf{Use Semantic Text for Progress Feedback}:
    The use of juicy semantic text should be considered against the use of particles. However, care must be taken in the choice of text content, since the feedback can also change the player's perception of their abilities.
 
 \textbf{Meaningful vs. Meaningless Text}: Performance in the MS hints at some amount of confusion when using gibberish, or meaningless, JT, but without much impact on the overall experience. Non-semantic JT could therefore be used, for example, to represent strange or indescribable sounds or elements.
 
 \textbf{Onomatopoeic vs. Semantic Text}: We can draw no firm conclusions here, on juiciness or otherwise.
    Each yields similar results to particles and the choice seems to be about what suits the game.
  
  \textbf{The Inherent Multimodality of Onomatopoeia}: In this study, we only focused on the visual aspect of onomatopoeia. However, games are rarely soundless. Future work on juicy and non-juicy onomatopoeic text effects should consider how they would fare when coupled with sound, perhaps especially juiced-up sound.

Text-based design offers a novel way for designers to convey information about the game world. Further research is needed to fully explore this modality, but our findings suggest a path towards more inclusive and engaging game design in the future.

\subsection{Limitations and Future Work}
Our game was simple, lacking, e.g., progression, a story, etc. Individual games may also have a unique way of conveying information that affects PXI. 
Comparing similar Onomatopoeic and Semantic effects (e.g., "BANG" vs. "EXPLOSION") would be insightful. Future analyses of performance should take user's varied skill levels and familiarity with the game type into account~\cite{DELMAS2022103628,ognjanovic2019display,CHISHOLM201593}.

Like particles~\cite{Kao2020}, different levels of juiciness across text effects could be investigated, considering the impact on visual clutter in game design~\cite{DELMAS2022103628}. 
We focused here mostly on the visual aspect of text effects akin to comic books. However, in games, text effects can be underwhelming without accompanying sound. Further research is needed to compare these effects in optimized, multimodal forms.


\section{Conclusion}
We have provided empirical evidence supporting the continued significance of particle effects in juicy effect design. Text effects have merits in conveying information and contributing to stylistic choices, but do not surpass particle effects in PX.
Our work explored an approach rooted in comic book aesthetics to create VFX infused with language. We shed light on the nuance between semantic and onomatopoeic text effects and a potential relationship with sound. \textbf{\textit{Although these text-based approaches to effect design may not supersede current practice, they serve as a valuable addition to the toolkit of VFX designers.}}
We offer a foundation for further studies of text effects and their capabilities. As we continue to navigate the landscape of juiciness and effect design, it is imperative to consider the importance of stylistic choices and environmental design in shaping the final interactive experience.

\begin{acks}
We thank the members of Rekimoto Lab for supporting this work. This work was partially supported by JST Moonshot R\&D Grant Number JPMJMS2012, JST ASPIRE Program Grant Number JPMJAP2327 and JST FOREST Grant Number JPMJFR206E.
\end{acks}

\bibliographystyle{ACM-Reference-Format}
\bibliography{references}




\end{document}